\documentclass[A4paper,11pt]{article}
\pdfoutput=1

\usepackage{latexsym}
\usepackage{epsfig}
\usepackage{verbatim}
\usepackage{multirow}
\usepackage{fancybox}
\usepackage{shadow}
\usepackage{cite}
\usepackage{amscd, amsmath, amssymb, empheq}
\usepackage{bm, dsfont}
\usepackage{graphicx}
\usepackage{newtxtext, newtxmath}
\usepackage{xcolor}
\usepackage{mdframed}
\mdfsetup{%
linecolor=white,
backgroundcolor=gray!20,}

\newtheorem{theorem}{\sc Theorem}[section]
\newtheorem{lemma}[theorem]{\sc Lemma}
\newtheorem{corollary}[theorem]{\sc Corollary}

\newcommand{\eps}{\varepsilon}

\newcommand{\proofend}{{\medskip\medskip}}
\newcommand{\proof}{{\noindent\em Proof. }}

\author{
 Bernard Chazelle
\thanks{Department of Computer Science,
       Princeton University, 
{\tt chazelle}@{\tt cs.princeton.edu }}
\and
Kritkorn Karntikoon
\thanks{Department of Computer Science,
       Princeton University, 
{\tt kritkorn}@{\tt cs.princeton.edu }}
}

\title{
Quick Relaxation in Collective Motion
\thanks{This work was supported in part by NSF grant
CCF-2006125.
}}

\vspace{.5cm}

\date{}

\begin{document} \maketitle

\begin{abstract}
We establish sufficient conditions for the quick
relaxation to kinetic equilibrium 
in the classic Vicsek-Cucker-Smale model of bird flocking.
The convergence time is polynomial in the number of birds
as long as the number of flocks remains bounded.
This new result relies on two key ingredients:  exploiting the
convex geometry of embedded averaging systems; and deriving
new bounds on the $s$-energy of disconnected agreement systems.
We also apply our techniques to bound the relaxation time of certain
pattern-formation robotic systems investigated by Sugihara and Suzuki.
\end{abstract}

\vspace{2cm}

\section{Introduction}

In the classic Vicsek-Cucker-Smale 
model~\cite{CuckerSmale1, vicsekCBCS95}, a group of $n$ birds are flying in 
the air while interacting via a time-varying 
network~\cite{blondelHOT05, chazFlockPaperI, HendrickxB, jadbabaieLM03}.
The vertices of the network correspond to the
$n$ birds and any two distinct birds are joined by an edge if their distance is at most some 
fixed $r\leq 1$.
The flocking network $G_t$ is thus
symmetric and loopless.  Its connected components are the {\em flocks}.
Each bird $i$ has a position $x_i(t)$ and a velocity $v_i(t)$,
both of them vectors in $\mathbb{R}^3$. Given the state of the system
at time $t=0$, we have the recurrence: for any $t\geq 0$, 

\begin{equation}\label{modelVCS}
\begin{cases}
\, x_i(t+1)= x_i(t)+ v_i(t+1); \\
\, v_i(t+1)= v_i(t) + a_i\sum_{j\in N_i(t)}\bigl( v_j(t) - v_i(t)\bigr),
\end{cases}
\end{equation}
where $N_i(t)$ is the set of vertices adjacent to $i$ at time $t$.
At each step, a bird adjusts its velocity
by taking a weighted average with its neighbors.
The weights $a_i$ 
indicate the amount of influence birds
exercise on their neighbors.  To avoid negative weights, we
require that $0<a_i\leq  1/(|N_i(t)|+1)$. We write $\rho:= \min_i a_i \in (0,1/2]$.
As was shown in~\cite{chazFlockPaperI}, the model above might be periodic
and never stabilize.  To remedy this, we stipulate that, for
two birds to be newly joined by an edge,  
their velocities must differ by at least a minimum amount:
Formally, we require that, at any time $t$, 
$(i,j)\in G_t\setminus G_{t-1}$ if $\|x_i(t)-x_j(t)\|\leq r$ 
and $\|v_i(t)- v_j(t)\| > \eps_o$, for small fixed positive~$\eps_o$.
By space and time scale invariance, we may assume\footnote{These
bounds are arbitrary and the choice of $\sqrt{\rho / n}$ is made only
to simplify some calculations.}
that 
$\|x_i(0)\| \leq 1$ and $\|v_i(0)\| \leq \sqrt{\rho / n}$ for all birds $i$.
We state our main result:

\vspace{0.4cm}
\begin{mdframed}
\begin{theorem}\label{ConvergeTime}
$\!\!\! .\,\,$
A group of $n$ birds forming a maximum of $m\leq n$ flocks
relax to within $\eps$ of a fixed velocity vector in 
time $O(n^2/\rho)\log (1/\eps) +t_o$, where
$t_o= O\bigl(n^2/\rho \bigr)^{m+2}\log (n/\rho)$.
\end{theorem}
\end{mdframed}
\vspace{0.4cm}

\vspace{0cm}
\begin{figure}[htb]
\begin{center}
\hspace{0cm}
\includegraphics[width=6cm]{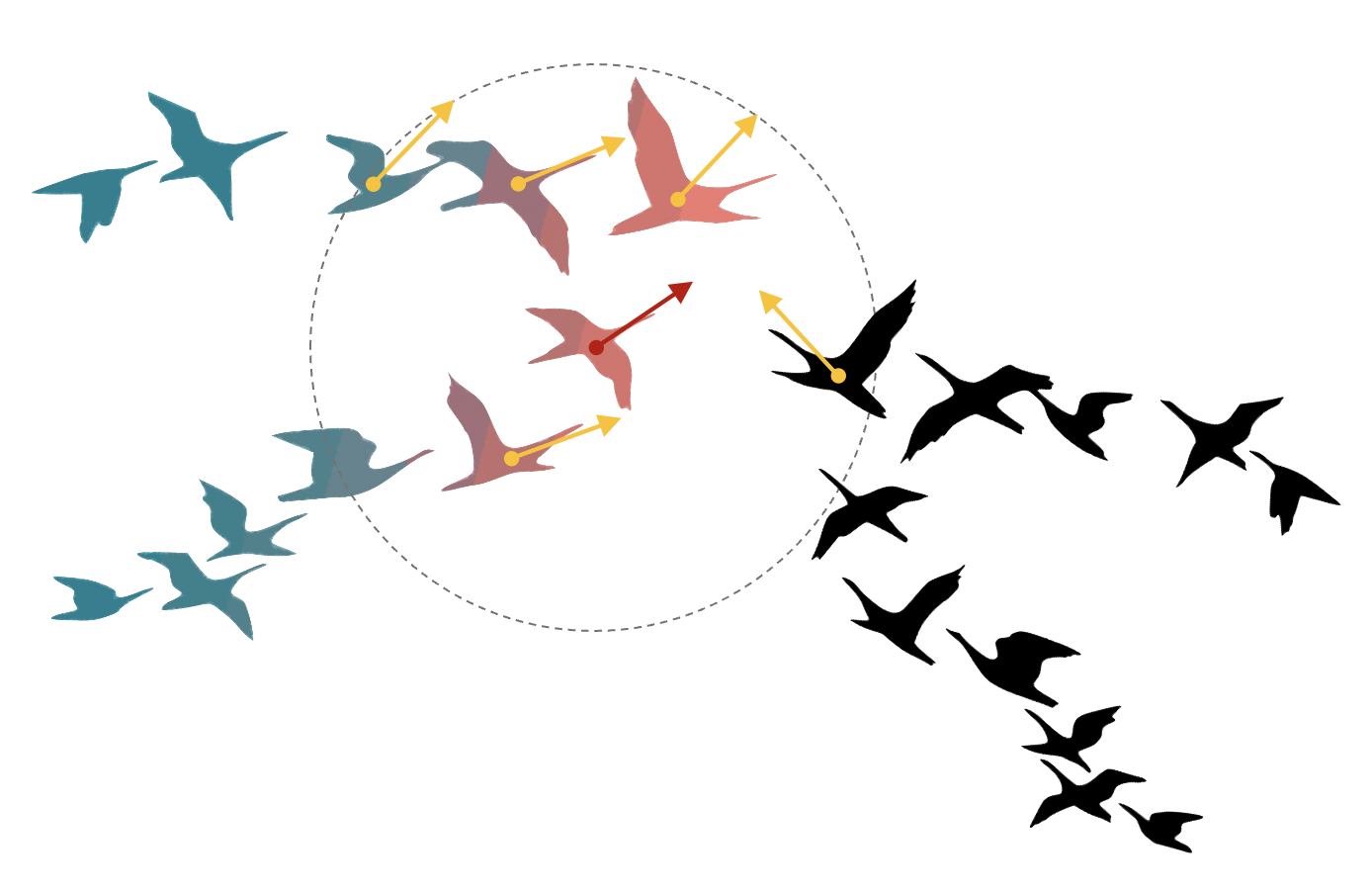}
\end{center}
\vspace{-0.5cm}
\caption{\small A bird is influenced by its neighbors within distance $r$ . \label{fig-flock}}
\end{figure}
\vspace{0.5cm}

The main novelty of this result is that
the convergence time is polynomial in the number of birds,
as long as the number of flocks is bounded by a constant.
The proof of the theorem relies on two key ingredients:
new bounds on the {\em $s$-energy}; and the specific convex geometry of flocking.
In~\S\ref{RAS}, we establish new (upper and lower) bounds on 
the $s$-energy of {\em reversible agreement systems}.
While the connected case has been well-studied~\cite{chazelle-total, chazelle-Energ2-2019},
the disconnected case was wide open. We prove a nearly tight upper bound 
on the $s$-energy of such systems, which is a result of independent interest.
In~\S\ref{flock}, we explore the convex geometry of flocking
to bound the angle of attack between two newly joined birds as
a function of time. Together with our new bounds on the $s$-energy,
this geometric insight plays 
a crucial part in the proof of Theorem~\ref{ConvergeTime}.

In ~\S\ref{PatternFormation}, we investigate a distributed motion coordination
algorithm introduced by~Sugihara and Suzuki~\cite{sugihara-suzuki-1990, bulloBk}.
The idea is to use a swarm of robots to produce a preset pattern, in this case
a polygon. We prove a polynomial bound on the relaxation time of this process.
We enhance the model by allowing faulty communication and proving
that the end result is robust under stochastic errors. We also generalize
the geometry to 3D and arbitrary communication graphs.

\section{Reversible Agreement Systems}\label{RAS}

Let $P_{t}$ be the stochastic matrix of a time-reversible random walk in 
an undirected $n$-vertex graph $G_t$.
This means that $P_t= Q^{-1} M_t$, where
(i) $Q= \hbox{diag}(q)$, for 
$q= M_t\mathbf{1}\preceq  \mathbf{1}/\rho$ and constant $\rho\in (0,1/2]$;
and (ii) $M_t$ is a symmetric matrix with
nonzero entries at least~1 and a positive diagonal.
Given $x\in \mathbb{R}^n$,
the infinite sequence of vectors $(P_{t} \cdots P_{ 0}x)_{t\geq 0}$
forms an orbit of a {\em reversible averaging system} ({\em RAS}).
When all the matrices are the same, $P_t= P$,
the map $x\mapsto Px$ is the dual map
of the reversible Markov chain ($y\mapsto yP$) and 
its convergence time is given by the mixing time of the chain.
The novelty of the model lies in the presence of time-varying matrices.

\subsection{The $s$-energy}\label{def-s-E}

Consider an infinite sequence of graphs $(G_t)_{|t\geq 0}$
with $P_t$ the stochastic matrix of its corresponding Markov chain.
Note that $q$ is proportional to the stationary distribution
of the Markov chain induced by $P_t$. By reversibility,
we have $q_i (P_t)_{ij}= q_j (P_t)_{ji}$.
Write 
$\langle x, y \rangle_q : = \sum_i q_i x_iy_i$
and $\|x\|_q^2:= \langle x, x \rangle_q$.
We call $\|x-  \hat x\, \|_q^2$ the {\em variance} of the system, where
$\hat x= \|q\|_1^{-1}\langle x, \mathbf{1}\rangle_q\mathbf{1}$
and $x$ is shorthand for $x(0)$.

Given $x(0)\in  \mathbb{R}^n$, we write $x(t+1)= P_t x(t)$
and we interpret $x(t)$ as an embedding of the graph $G_t$ in $\mathbb{R}$.
The union of the embedded edges of $G_t$ forms disjoint
intervals, called {\em blocks}.
Let $l_1,\ldots, l_k$ be the lengths of these blocks
and put $E_{s,t}= \sum_{i=1}^k l_i^s$, with $s\in (0,1]$.
Following~\cite{chazelle-Energ2-2019} ,
we define the {\em $s$-energy} $E_s= \sum_{t\geq 0} E_{s,t}$.
We denote by $\mathcal{E}_{m,s}$ the supremum 
of $E_s$ over all systems of unit variance
whose underlying graphs $G_t$ have at most $m$ connected components.

\subsection{An upper bound}

We prove the following bound on the $s$-energy of reversible
agreement systems and then we show in~\S\ref{LB-sEn} why it is close to optimal.

\vspace{0.3cm}
\begin{mdframed}
\begin{theorem}\label{reversible-bound}
$\!\!\! .\,\,$
\  $\mathcal{E}_{m,s}
\leq (cn^2/\rho s)^m$, for any $s\in (0,1]$
and constant $c>0$.
\end{theorem}
\end{mdframed}
\vspace{0.3cm}

\proof
The convergence rate of the attracting dynamics
is captured by a variant of the Dirichlet form:
$D_t= \sum_{i}\max_{j:\, (i,j)\in G_t} \bigl( x_i(t)- x_j(t) \bigr)^2$.
We omit the index $t$ below for clarity.

\begin{lemma}\label{dirich}
$\!\!\! .\,\,$
\ 
$\|Px\|_q^2  \leq \|x\|_q^2 - D/2\,$, for any $x\in \mathbb{R}^n$.
\end{lemma}
\proof
Write $\delta_{ij}= x_i-x_j$ and $\mu_i = \sum_{j} p_{ij}  \, \delta_{ij}$.
Fix $i$ and pick any $k$ such that $m_{ik}>0$.
By Cauchy-Schwarz and $m_{ii}\geq 1$, we have
$$
\delta_{ik}^2= \bigl( (\mu_i- \delta_{ii}) + (\delta_{ik}-\mu_i) \bigr)^2
   \leq 2(\delta_{ii}-\mu_i)^2 + 2(\delta_{ik}-\mu_i)^2
   \leq 2\sum_{j} m_{ij}  \, (\delta_{ij}- \mu_i)^2;
$$
hence,
\begin{equation*}
\begin{split}
\|x\|_q^2- \|Px\|_q^2  
&=  \sum_i q_i x_i^2 -  \sum_i q_i \,\biggl(x_i+ \sum_j p_{ij}\delta_{ji}\biggr)^2
= - \sum_i q_i \, \biggl( 2 x_i \sum_j p_{ij}\delta_{ji} +  \mu_i^2  \biggr) \\
&=  \sum_i q_i \, \biggl( \, \sum_{j} p_{ij}  \, \delta_{ij}^2   -  \mu_i^2 \, \biggr) 
=  \sum_{i,j} m_{ij}  \, (\delta_{ij}- \mu_i)^2 
\geq \frac{1}{2}\sum_{i}  \max_{j: m_{ij}>0} \delta_{ij}^2,
\end{split}
\end{equation*}
with the last equality expressing the identity for the variance:
$\mathbb{E} X^2 - (\mathbb{E}\, X)^2
= \mathbb{E}[X- \mathbb{E}\,X]^2$.
\hfill $\Box$
\proofend

Write $G_{\leq t}$ as the union of all the edges in $G_0,\ldots, G_t$,
and let $t_c$ be the maximum value of $t$ such that
$G_{\leq t}$ has fewer connected components than $G_{\leq t-1}$;
if no such $t$ exists, set $t_c=1$.

\begin{lemma}\label{cover-length}
$\!\!\! .\,\,$
If $G_{\leq t_c}$ is connected, then
$\sum_{t\leq t_c} D_t \geq \, \rho n^{-2}\,  \|x-\hat x\, \|_q^2\,$.
\end{lemma}
\proof
Let $G_{t_0}$ denote the graph over $n$ vertices with no edges.
We define $t_1,\ldots, t_c$ as the sequence of times $t$ at which
the addition of $G_{t}$ reduces the number of connected components in $G_{\leq t-1}$.
At any time $t_k$ ($k>0$), 
the drop $d_k$ in the number of components can be achieved by $d_k$
edges from $G_{t_k}$.
Let $F_k$ denote such a set of edges: we can always
order $F_k$ so that every edge in the sequence
contains at least one vertex not encountered yet.
This shows that the sum of the squared lengths
of the edges in $F_k$ does not exceed $D_{t_k}$.
We note that 
$F:= F_1\cup\cdots\cup F_c$ forms a collection of $n-1$ edges
from (the connected graph) 
$G_{\leq t_c}$ and $F$ spans all $n$ vertices.

Consider the intervals formed by the edges in $F_k$ at time $t_k$,
for all $k\in [c]$. 
The union of these intervals covers the smallest interval $[a,b]$
enclosing all the vertices at time $0$ (and hence at all times). 
To see why, pick any $z$ such that $a<z<b$ and denote by $L$
and $R$ the vertices on both sides of $z$ at time $0$.   Neither set is empty and,
by convexity, both of them remain on their respective side
of $z$ until an edge of some $G_t$ joins $L$ to $R$.
When that happens (which it must since $G_{\leq t_c}$ is connected), the joining edge(s)
reduce(s) the number of components of $G_{\leq t-1}$ 
by at least one, so $F$ must grab at least one of them,
which proves our claim. Let $l_1,\ldots, l_{n-1}$ denote the
lengths of the edges of $F$ (at the time of their insertion).
By Cauchy-Schwarz, 
$\sum_{t\leq t_c} D_t \geq \sum_{i=1}^{n-1} l_i^2 \geq (b-a)^2/(n-1)$.
The lemma follows from the inequalities
$\|x-\hat x\, \|_q^2\leq \|q\|_1(b-a)^2\leq (n/\rho) (b-a)^2$.
\hfill $\Box$
\proofend

\smallskip

Assume that $G_{\leq t_c}$ is connected.
By Lemma~\ref{cover-length} and the telescoping use of Lemma~\ref{dirich},
\begin{equation}\label{telescope}
 \|x\|_q^2 -  \|x(t_c+1)\|_q^2
\geq \frac{1}{2} \sum_{t=0}^{t_c} D_t
\geq \frac{\rho}{2n^2} \, \|x-\hat x\, \|_q^2 \, .
\end{equation}

\noindent
Let $U(n,m)$ be the maximum $s$-energy of an {\em RAS}
with at most $n$ vertices and $m$ connected components at any time,
subject to the initial condition $\|x - \hat x\, \|_q^2 \leq 1$.
By shifting the system if need be, we can always assume that $\hat x= \mathbf{0}$.
By~(\ref{telescope}),
$\|x(t)\|_q^2$ shrinks by at least a factor of $\alpha:= 1- \rho/2n^2$ by time $t_c+1$.
A simple scaling argument shows that
the $s$-energy expanded after $t_c$ is at most $\alpha^{s/2} U(n,m)$.
While $t<t_c$ (or if $G_{\leq t_c}$ is not connected),
the system can be decoupled into two {\em RAS},
each one with fewer than $m$ components.\footnote{Note that
each subsystem satisfies the required
inequalities about the $Q$ and $M$ entries; also, shifting
each subsystem so that $\hat x=0$ cannot increase
$\|x\|_q^2$, so its value remains at most 1.}
Since $\|x\|_q^2 \leq 1$, 
the diameter at any time is at most 2; therefore
$U(n,m) \leq \alpha^{s/2} U(n,m) + 2 U(n,m-1) + m2^s$.
It follows that 
\begin{equation}\label{U-recur}
U(n,m) \leq \frac{2}{1-\alpha^{s/2}}\, \bigl( U(n,m-1) + m \, \bigr);
\end{equation}
hence $U(n,m) = O(n^2/\rho s)^m$
and the proof of Theorem~\ref{reversible-bound} is complete.
\hfill $\Box$
\proofend

\subsection{A lower bound}\label{LB-sEn}

We begin with the case $m=1$.
The path graph $G$ over $n$ vertices has an edge $(i,i+1)$ for all $i<n$.
The Laplacian $L$ is $\text{diag} (u) -A$, where $A$ is the adjacency matrix of $G$
and $u$ is the degree vector $(1,2,\ldots, 2,1)$. We consider 
the {\em RAS} formed by the matrix $P_t= P= I - \rho L$.
By well-known spectral results on graphs~\cite{spielman19},
$P_t$ has a full set of $n$ orthogonal eigenvectors
$v_k$, where $v_k(i)=  \cos \frac{(i-1/2)k\pi}{n}$ for $i\in [n]$,
with its associated eigenvectors 
$\lambda_k= 1- 2\rho\bigl(1-  \cos \frac{ k\pi}{n}\bigr) $, for $0\leq k<n$.
We require $\rho<1/4$ to ensure that $P$ is positive semidefinite.
We initialize the system with $x= (1,0,\ldots, 0)$ and observe that
the agents always keep their initial rank order, so the diameter $\Delta_{t}$ at time $t$
is equal to $(1,0,\ldots, 0,-1)P^{t-1} x$.
We verify that $\|v_1\|^2= n/2$. By the spectral identity 
$P^j= \sum_{k<n} \lambda_k^j  v_k v_k^T/  \|v_k\|^2$, 
we find that, for $t>1$ and $n>1$,
\begin{equation*}
\Delta_t
= \sum_{k=0}^{n-1} \lambda_k^{t-1}  v_k(1)\frac{v_k(1)-v_k(n)}{\|v_k\|^2}
= \sum_{{\rm\small odd}\,\,\, k}
\frac{2\lambda_k^{t-1}}{\|v_k\|^2} \Bigl(\cos\frac{k\pi}{2n} \Bigr)^2
\geq  \frac{2\lambda_1^{t-1}}{\|v_1\|^2} \Bigl(\cos\frac{\pi}{2n} \Bigr)^2
\geq \frac{2}{n}\lambda_1^{t-1} .
\end{equation*}
The $s$ energy is equal 
to $\sum_t \Delta_t^s\geq  (2/n)^s/(1-\lambda_1^s)
\geq bn^{2-s}/\rho s$, for constant $b>0$.

For the general case, we denote by $F(n, m)$ the $s$-energy of the 
system with initial diameter equal to 1. We showed that
$F(n,1)\geq  bn^{2-s}/\rho s$.
We now describe the steps of the dynamics for $m>1$.
To simplify the notation, we assume that
$\nu:= n/m$ is an integer.\footnote{This can be relaxed with a simple
padding argument we may omit.}
For $i\in [m]$, let $C_i$ be the path linking vertices $[(i-1)\nu +1, i\nu]$.

\begin{enumerate}
\item
At time $t=1$, the vertices of $C_1$ are placed at position $0$
while all the others are stationed at $1$. 
The paths $C_1$ and $C_2$ are linked together into a single path
so the system has $m-1$ components. Vertices $\nu$ and $\nu+1$ move
to positions $\rho$ and $1-\rho$ respectively while the others do not move at all.
The $s$-energy expended during that step is equal to 1.
\item
The system now consists of the $m$ paths $C_i$.  We apply the case $m=1$
to $C_1$ and $C_2$ in parallel, which expends $s$-energy
equal to $2\rho^s F(\nu, 1)$. All other vertices stay in place.
The transformation keeps the mass center invariant, so 
the vertices of $C_1$ and $C_2$ end up at positions $\rho/\nu$
and $1- \rho/\nu$, respectively.\footnote{To keep the time finite,
we can always force completion in a single step
once the agents are sufficiently close to each other
and use a limiting argument.} 
\item
We move the vertices in $C_i$ for $i\geq 2$ by applying
the same construction recursively for fewer than $m$ components.
The vertices of $C_1$ stay in place.
The $s$-energy used in the process is equal to 
$\bigl(\rho/\nu\bigr)^s F(n-\nu, m-1)$ 
and the vertices of $C_2,\ldots, C_m$
end up at clustered at position $1- \frac{\rho}{n-\nu}$.
\item
We apply the construction recursively to the $n$ vertices,
which uses up a quantity of $s$-energy equal to 
$\bigl( 1-\frac{\rho}{\nu} - \frac{\rho}{n-\nu} \bigr)^s F(n,m)$.
\end{enumerate}
Putting all the energetic contributions together, we find that,
for constants $b', c>0$,
\begin{equation*}
\begin{split}
F(n,m) &\geq 
1 +
\frac{2b \nu^{2-s}}{s \rho^{1-s}}  + 
\left( \frac{\rho}{\nu} \right)^s F(n-\nu, m-1) +
\left( 1-\frac{2\rho}{\nu}  \right)^s F(n,m) \\
&\geq \frac{b'}{s\rho^{1-s}}\left(\frac{n}{m}\right)^{1-s}F(n-\nu, m-1)
\geq \left( \frac{c}{s\rho^{1-s}}\right)^m
\left(\frac{n}{m}\right)^{(1-s)m+1}.
\end{split}
\end{equation*}

\vspace{0.3cm}
\begin{mdframed}
\begin{theorem}\label{s-energy-RS-LB}
$\!\!\! .\,\,$
There exist reversible agreement systems with initial diameter equal to 1
whose $s$-energy is at least
$\bigl( c/s\rho^{1-s} \bigr)^m
(n/m)^{(1-s)m+1}$,
for constant $c>0$.
The number of vertices is $n$ and the number of connected
components is bounded by $m$; furthermore, all positive entries
in the stochastic matrices are at least $\rho$.
\end{theorem}
\end{mdframed}
\vspace{0.3cm}

Our lower bound constructions assume a unit diameter at time $0$.
Since $\mathcal{E}_{m,s}$ is defined for unit variance systems,
we must scale the bound appropriately to compare the lower
bound with Theorem~\ref{reversible-bound}.
We have $q= \mathbf{1}/\rho$,
so the variance is most $n/\rho$ and we need to scale the lower bound by
$(\rho/n)^{s/2}$.

\vspace{0.5cm}
\section{The Convergence Rate of Flocking}\label{flock}

We rewrite the map of the velocity dynamics~(\ref{modelVCS}) in matrix form,
$v(t+1)= P_t v(t)$, for $t\geq 0$, where $v(t)$ is an $n$-by-3 matrix with each row indicating
a velocity vector. We have
$P_t= Q^{-1}M_t$, where:
$Q= \hbox{diag}(q)$;
$q_i= 1/a_i$;
$(M_t)_{ij}= 1/a_i- |N_i(t)|$ if $i=j$ and $1$ else.
Note that $\bar q:= \|q\|_1^{-1} q$ is the joint stationary distribution
and $q= M_t\mathbf{1}\leq  \mathbf{1}/\rho$, where
$\rho:= \min_i a_i \in (0,1/2]$.
This shows that each one of the three coordinates provides
its own reversible agreement system $\mathcal{S}_j$ ($j=1,2,3$).
The only difference between the systems is their initial states.
Recall that the $s$-energy of any such system is defined as
$\sum_t E_{s,t}$, where $E_{s,t}= \sum_i l_i(t)^s$
and $l_i(t)$ is the length of the $i$-th block at time~$t$.
Let $m$ be the maximum number of flocks and
$N_{m,\alpha}$ the number of times~$t$ at which
some block length $l_i(t)$ from at least one of 
$\mathcal{S}_j$ ($j=1, 2, 3$) exceeds $\alpha$.
For $0<\alpha<1$, we have 
$N_{m,\alpha}\leq \inf_{s\in (0,1]} 3 \alpha^{-s} \mathcal{E}_{m,s}$.
Our assumption that $\|v_i(0)\|^2\leq \rho / n$ for all birds $i$
implies that the three systems have variance at most one.
By Theorem~\ref{reversible-bound}, for some (other) constant $c>0$,
setting $s= 1/ \log (1/\alpha)$ yields 

\noindent
\begin{equation}\label{N-alpha}
N_{m,\alpha} \leq  \left(\frac{cn^2}{\rho}\log \frac{1}{\alpha} \right)^m .
\end{equation}

\subsection{Single-flock dynamics}\label{SF-dyn}

Between two consecutive switches (ie, edge changes), the flocking networks consists
of fixed non-interacting flocks. We can analyze them separately.
Without loss of generality, assume that $G_t$ is a connected, time-invariant graph.
We focus on system $\mathcal{S}_1$ for convenience.
It consists of a single block at each timestep, so the $s$-energy is
of the form $\sum_t  \Delta_t^s$, where $\Delta_t$ is the diameter of
the system at time $t\geq 0$. 
The diameter can never grow, so by
the same argument leading to~(\ref{N-alpha}),
we know that $\Delta_t \leq \alpha$ for any 
$t\geq  \inf_{s\in (0,1]} \alpha^{-s} \mathcal{E}_{1,s}$.
It follows that $\Delta_t\leq e^{-a \rho t/n^2}$, for constant $a>0$.
Recall that $x(t)$ and $v(t)$ are $n$-by-3 matrices; 
denote their first column by $y(t)$ and $w(t)$, respectively.
Write $y(0)= y$, $w(0)=w$, and $P_t=P$.
The vector $w(t) = P^t w$ tends to $(\bar q^T w) \mathbf{1}$.
Since its coordinates lie in an interval of width $\Delta_t$, it follows that
$w(t)= (\bar q^T w) \mathbf{1} + \zeta(t)$, where
$\| \zeta(t) \|_\infty \leq \Delta_t\leq  e^{- a\rho t/n^2}$.
Thus, for some $\gamma , \eta_t \in \mathbb{R}^n$,
$$
y(t) = y + \sum_{k=1}^{t} w(k)= y +  
t (\bar q^T w) \mathbf{1} + \sum_{k=1}^t \zeta(k)
= \beta t + \gamma + \eta_t,
$$
where
$\beta= ( \bar q^T w ) \mathbf{1}$,
$\gamma= y+ \sum_{k=1}^\infty \zeta(k)$,
and $\| \eta_t \|_\infty\leq e^{-b\rho t/n^2}$, for constant $b>0$.
The same holds true of the other two coordinates, so
the birds in the flock fly parallel to a straight line with a deviation
from their asymptotic line vanishing exponentially fast.
If so desired, it is straightforward to lock the flocks by stipulating that 
no two birds can lose an edge between them unless their velocities
exceed a small threshold~$\theta$; because of the 
exponential convergence rate, choosing $\theta$ small enough
ensures that two birds $i$ and $j$ adjacent in a flock may exceed
distance~$r$ by only a tiny amount.

\subsection{Flock fusion}

To bound the relaxation time, we begin with an
intriguing geometric fact:
Far enough into the future, two birds can only come close
to each other if their velocities are nearly identical.
In other words, encounters at large angles of attack 
cannot occur over a long time horizon.
We begin with a technical lemma: A stationary
observer positioned at the initial location of a bird
sees that bird move less and less over time;
this is because the bird flies increasingly in the direction of the line of sight.

\begin{lemma}\label{line-of-sight}
$\!\!\! .\,\,$
For constant $c$ and any $t>1$,
$\left\| v_i(t) - \frac{1}{t} \bigl(x_i(t) - x_i(0) \bigr) \right\|
\leq 
\bigl( c n^2/\rho \bigr)^{m+2} (\log t)/t$,
for any bird~$i$.
\end{lemma}
\proof
For notational convenience, we set $i=1$ and
we denote by $y_j(t)$ (resp. $w_j(t)$)
the first coordinate of $x_j(t)$ (resp. $v_j(t)$).
The line-of-sight direction of bird 1 is given by $\frac{1}{t}\bigl( x_1(t)-x_1(0) \bigr)$.
Along the first coordinate axis, this gives us
\begin{equation}\label{u=sum}
u:= \frac{1}{t}\bigl( y_1(t) - y _1(0) \bigr)
 =  \frac{1}{t} \sum _{k=1}^{t} w_1(k).
\end{equation}
Consider the difference $\delta:= u - w_1(t)$. We can
define the corresponding quantity for each of the other two directions
and assume that $\delta$ has the largest absolute value among the three of them.
By symmetry, we can also assume that $\delta\geq 0$; therefore
\begin{equation}\label{u-diff}
\bigl\| v_1(t) - \hbox{$\frac{1}{t}$} \bigl(x_1(t)- x_1(0) \bigr) \bigr\|
\leq \sqrt{3}\, \delta. 
\end{equation}

The proof of the lemma rests on showing that, if $\delta$
is too large, some bird $l$ must be at a distance greater than 1
from bird $1$ at time 0, which has been ruled out. To identify the far-away bird~$l$,
we start with $l=1$ at time $t$, and we trace
the evolution of its flock backwards in time, always trying to move
away from bird~1, if necessary by switching bird $l$ with a neighbor.
This is possible because of two properties, at least one of which holds
at any time $k$:  (i) bird~$l$ flies nearly straight in the time interval $[k, k+1]$;
or (ii) bird~$l$ is adjacent to a bird~$l'$ whose velocity points in a favorable
direction. In the latter case, we switch our focus from $l$ to $l'$.

The $s$-energy plays the key role in putting numbers behind these 
properties. For this reason, we define $\mu_l(k)$ as the length of
the block of $\mathcal{S}_1$ containing $w_l(k)$ with respect to 
the flocking network $G_k$. 
Note that $\mu_l(k)$ is the length of an 
interval that contains the numbers $w_j(k)$ for all the birds $j$
in the flock of bird $l$ at time~$k$.
We define the sequence of velocities $\bar{w}(k) = w_l(k)$,
for $k= t, t-1,\ldots, 1$ and $l= l(k)$.
Fix some small $\alpha$ ($0<\alpha \leq \eps_o$).

\vspace{0.5cm}

{\small
\par\medskip
\renewcommand{\sboxsep}{0.5cm}
\renewcommand{\sdim}{0.8\fboxsep}
\begin{center}
\shabox{\parbox{10cm}{
\begin{itemize}
\item[\text{[1]}]
\hspace{0.2cm}
$\bar{w}(t) \leftarrow w_1(t)$ \ \ and \ \ $l \leftarrow 1$
\item[\text{[2]}]
\hspace{0.2cm}
{\bf for }\  $k=t-1, \ldots, 1$
\item[\text{[3]}]
\hspace{0.8cm}
{\bf if } \ $\mu_l(k) > \alpha$
\ {\bf then } \ $l\leftarrow \text{argmin} \left\{\, w_j(k) \, |\,  j\in N_l(k)\,  \right\}$
\item[\text{[4]}]
\hspace{0.8cm}
$\bar{w}(k) \leftarrow w_l(k)$
\end{itemize}
}}
\end{center}
\par
}
\vspace{0.8cm}
Perhaps the best way to understand the algorithm is first to imagine
that the conditional in step [3] never holds:
In that case, $l=1$ throughout and we are simply
tracing the backward evolution of bird 1.  Step [3] aims to catch
the instances where the reverse trajectory inches excessively toward
the initial position of bird 1.
When that happens, $|w_l(k+1)- w_l(k)|$ is large, hence so
is $\mu_l(k)$, and step [3] kicks in.
We exploit the fact that $w_l(k+1)$ is a convex
combination of $\bigl\{w_j(k) \, |\,  j\in N_l(k)\bigr\}$ to update
the current bird $l$ to a ``better" one.
Using summation by parts, we find that
\begin{equation}\label{sum-parts}
 \sum_{k=1}^{t} \bar{w}(k) =
 t \bar w(t) -
 \sum_{k=1}^{t-1} k\bigl( \bar{w}(k+1) - \bar{w}(k)\bigr).
 \end{equation}
 
Let $R$ be the set of times $k$ that pass the test in step [3]
and $S$ the set of switches (ie, network changes).
An edge creation entails a block of length $\eps_o/\sqrt{3}$ or more
in at least one of $\mathcal{S}_j$ ($j=1,2,3$).
The steps witnessing edge deletions outnumber those 
seeing edge creations by at most a factor of $\binom{n}{2}$.
Let $I$ be the time interval between two consecutive switches.
Each flock remains invariant during $I$; therefore $|R\cap I|\leq N_{1,\alpha}$;
hence 
\begin{equation}\label{SR}
|S|\leq  n^2 N_{m, \eps_o/\sqrt{3}} 
\hspace{0.7cm} 
\text{and}
\hspace{0.7cm}
|R|\leq  (N_{1,\alpha} +1) |S|.
\end{equation}
\noindent
Because of the single-flock invariance, the diameter of $\mathcal{S}_1$ during $I$ 
can never increase; therefore $J= I\setminus R$ consists of a single time interval.
If $k\in J$, then  
$| \bar{w}(k+1) - \bar{w}(k) |
= |w_l(k+1) - w_l(k) | \leq \mu_l(k)\leq \alpha$ and, 
by~Theorem~\ref{reversible-bound},
$
\sum_{k\in J}  | \bar{w}(k+1) - \bar{w}(k) |
\leq \sum_{k\in J} E_{1,k}\leq \alpha \, \mathcal{E}_{1,1}= O(\alpha n^2/\rho)$;
hence 

\begin{equation}\label{Ek-alpha}
\sum_{k\in \{1,\ldots, t-1\}\setminus R}  \bigl| \bar{w}(k+1) - \bar{w}(k)  \bigr| =
O(\alpha n^2 |S| /\rho).
\end{equation}

\noindent
Let $l'$ be the value of $l$ in the final assignment
$\bar{w}_l(1) \leftarrow w_l(1)$ in step [4].
Since $\bar{w}(k+1) \geq \bar{w}(k)$ for $k\in R$
and $\bar{w}(t)= w_1(t)$, it follows 
from~(\ref{sum-parts}, \ref{Ek-alpha}) and $r\leq 1$ that
\begin{equation}\label{ylo-y}
\begin{split}
y_{l'}(0) - y_1(0)
&\geq  
\bigl(y_1(t) - y_1(0)\bigr) + \bigl(y_{l'}(0) -y_1(t) \bigr) 
\geq  tu -  \sum_{k=1}^{t} \bar{w}(k) - r |R|  \\
& \geq  t \delta 
+  \sum_{k=1}^{t-1} k\bigl( \bar{w}(k+1) - \bar{w}(k)\bigr)
      - |R| 
 \geq  t \delta -  O(t \alpha n^2 |S| /\rho)  -  |R|.
\end{split}
\end{equation}

\noindent
We set $\alpha= \eps_o /t$.
Noting that $y_{l'}(0) - y_1(0) \leq 1$, 
the lemma follows from~(\ref{N-alpha}, \ref{u-diff}, \ref{SR}) and
$$
\delta\leq 
n^{2(m+2)}
\left( \frac{b}{\rho}\right)^{m+1}
\left( \log \frac{1}{\eps_o} \right)^m 
\frac{\log (t/\eps_o)}{t} \, .$$
for constant $b>0$.
\hfill $\Box$
\proofend

\subsection{Stabilization}

By Lemma~\ref{line-of-sight}, for a large enough constant $c= c(\eps_o)$,
after time $t> t_o:= \bigl( c n^2/\rho \bigr)^{m+2}\log (n/\rho)$, 
no bird's velocity differs from its line-of-sight vector
$u_i= \frac{1}{t}\bigl( x_i(t) - x_i(0) \bigr)$ by a vector longer than $\eps_o /3$.
Suppose that birds $i$ and $j$ are within distance $r$ of each other.
By the triangular inequality, 
$ \|u_i - u_j\| \leq \frac{1}{t}\| x_i(t) - x_j(t)\| + \frac{1}{t} \|x_i(0) - x_j(0)\|\leq (1+r)/t$;
therefore,
$$
\|v_i(t)- v_j(t)\|\leq \|v_i(t)- u_i\| + \|u_i - u_j\| +  \|v_j(t)- u_j\| \leq \eps_o.
$$
This implies that each flock is time-invariant past time $t_o$.
The birds within each flock align their velocities exponentially fast.
In view of~\S\ref{SF-dyn}, this
completes the proof of Theorem~\ref{ConvergeTime}.
\hfill $\Box$
\proofend

\section{Distributed Motion Coordination}\label{PatternFormation}

In \cite{sugihara-suzuki-1990} Sugihara and Suzuki introduced 
an interesting model of pattern formation in a swarm of robots. In their model,
the robots can communicate anonymously and adjust their positions
accordingly. Assume that their goal is to align themselves along
a line segment $ab$. Two robots position themselves manually at the endpoint 
the segment while the others attempt to reach $ab$ by linking with their right/left
neighbors and averaging their positions iteratively.   This setup creates
a polygonal line $u_1=a, u_2,\ldots, u_{n-1}, u_n=b$, where $u_i$ is
the position $(x_i,y_i)$ of robot~$i$. The polygonal line converges to $ab$
in the limit. We use the $s$-energy bounds
to evaluate the convergence time of the robots. We actually prove 
a stronger result by generalizing the model in two ways:
(i) we consider the case of an arbitrary communication 
network of robots in 3D, with a subset of vertices pinned to a fixed plane;
(ii) the network suffers from stochastic edge failures.
Our model trivially reduces to Sugihara and Suzuki's by projection.
Allowing stochastic failures to their motion coordination model is novel. 

Let $G$ be a connected (undirected) graph with $n$ vertices labeled in $[n]$;
and let  the {\em communication weights} $a_1,\ldots, a_n$ be $n$ positive reals
such that $a_i<1/(d_i+1)$,
where $d_i$ is the degree of vertex $i$. We define
$d= \max d_i$ and $\rho= \min a_i$.
For any $t\geq 0$, we define $G_t$ by deleting each edge of $G$ with
probability $1-p$.
We define a (random) stochastic matrix $P_t$ for $G_t$ as follows:
\begin{quote}
\begin{enumerate}
\item
Initialize $P_t=0$;
\item
If $(i,j)$ is an edge of $G_t$, we set $(P_t)_{ij} = a_i$ and $(P_t)_{ji}=a_j$.
\item
$(P_t)_{ii}= 1- \sum_{j (j\neq i)} (P_t)_{ij}$, for all $i$.
\end{enumerate}
\end{quote}
Note that every positive entry of $P_t$ is at least $\rho$.
We embed $G$ in $\mathbb{R}^3$ and pin a subset $R$ of $r$ vertices
to a fixed plane. We fix the scale by assuming that
the embedding lies the 
unit cube $[0,1]^n$. Without loss of generality, we choose the plane $X=0$.
To ensure the immobility of the $r$ vertices, we can set $a_i=0$ for each $i\in R$.
Equivalently, we use {\em symmetrization}~\cite{chazelle-total} 
by attaching to $R$ a copy 
of $G$ and initializing the embedding of the two copies as mirror-image
reflections about $X=0$; note that the resulting
graph has $\nu= 2n-r$ vertices. 
The sequence $(P_t)_{t\geq 0}$ is defined by picking a random $G_t$ (as defined above)
at each step~{\rm iid}.

The vertices of $R$ are embedded in the plane $X=0$ at time 0, where,
by symmetry, they reside permanently. To prove the convergence of the
$\nu$ points to the plane, it suffices to focus on the dynamics along
the $X$-axis.  Given $x(0)\in [-1,1]^\nu$, we have
$x(t+1) = P_t x(t)$. This gives us a reversible agreement system.
Using the notation from~\S\ref{RAS} 
and Lemma~\ref{dirich}, we have $q= (1/a_1,\ldots, 1/a_\nu)$.
Since $G$ is connected, there is a path $\pi$ connecting the leftmost
to the rightmost vertex along the $X$-axis.
By the Cauchy-Schwarz inequality,

\begin{equation*}
\mathbb{E} \, D_t
\geq \mathbb{E} \sum_{i=1}^\nu  \max_{j:(i,j)\in G_t}\delta_{ij}^2
\geq   \sum_{i=1}^\nu \sum_{j:(i,j)\in G} p \delta_{ij}^2/d_i 
\geq \frac{p}{d\nu} \Bigl(\sum_{(i,j)\in \pi} |\delta_{ij}| \Bigr)^2
\geq  \frac{\rho p}{d\nu^2} \|x\|_q^2 \, ,
\end{equation*}

\noindent
where $\delta_{ij} = \delta_{ij}(t) = x_i(t) - x_j(t)$. It follows from Lemma~\ref{dirich} that,
for $c:= \rho p / (2d \nu^2)$,
$$
\mathbb{E} \, \|Px\|_q^2  \leq \|x\|_q^2 - \frac{1}{2}\,  \mathbb{E} \, D_t
\leq ( 1 - c ) \|x\|_q^2.
$$
By Markov's inequality,
$$
\Pr\left[\|Px\|_q^2 \geq \Bigl(1-\frac{c}{3}\Bigr) \|x\|_q^2\right] 
\leq \frac{\mathbb{E}\,  \|Px\|_q^2  }{ (1 - c/3)  \|x\|_q^2}\leq
1-\frac{c}{2} \, .
$$

\noindent 
Let $l_1,\ldots, l_k$ be the lengths of the blocks
formed by the edges of $G_t$ embedded along 
the $X$-axis.\footnote{Recall that the blocks are the intervals formed
by the union of the embedded edges of $G_t$.}
In a slight variant, we define the $s$-energy $E_s= \sum_{t\geq 0} E_{s,t}$,
where $E_{s,t}= \max_{i=1}^k l_i^s$.
We denote by $W_s$ the maximum expected 
$s$-energy, where the maximum is taken over all initial positions
with variance $\|x\|_q^2 \leq \nu/\rho$ 
(see~\S\ref{def-s-E} for  definitions). 
Since the vertices are embedded in $[-1,1]$ with symmetry about the origin,
this applies to the case at hand.
By Cauchy-Schwarz, we see that the diameter is at most
$\sqrt{2}\, \|x\|_q \leq \sqrt{2\nu/\rho}$. 
By scaling invariance, we have the following recurrence relation:
\begin{equation*}
\begin{split}
W_s &\leq \left(\frac{2\nu}{\rho}\right)^{s/2} + 
\frac{c}{2}\left(1 - \frac{c}{3}\right)^{s/2}W_s
+ \left(1 - \frac{c}{2}\right)W_s\\
&\leq  \frac{2^{s+1}\nu^{s/2}}{c\rho^{s/2}\big(1 - (1-c/3)^{s/2}\big)}
= O\left(\frac{\nu^{s/2}}{sc^2\rho^{s/2}}\right)
=  O\left( \frac{d n^2}{\rho p}\right)^2 \frac{(n/\rho)^{s/2}}{s} \, .
\end{split}
\end{equation*}
Let $N_\alpha$ be the number of times $t$ at which some block length $l_i(t)$ exceeds $\alpha$. 
For $0 < \alpha < 1$, we have $\mathbb{E}\, N_\alpha  \leq \inf_{s \in (0,1]} \alpha^{-s}W_s$. 
Setting $s = 1/\log (n/\rho \alpha^2)$ yields 
$$
\mathbb{E}\,N_\alpha =  
O\left( \frac{d n^2}{\rho p}\right)^2 \log\frac{n}{\rho\alpha}\, .
$$
Let $K_\alpha$ be the number of times $t$ at which there exists an edge $(i,j) \in G_t$ 
whose length $|\delta_{ij}(t)|$ exceeds $\alpha$;
obviously, $K_\alpha \leq N_\alpha$.
Let $T_\alpha$ be the last time at which 
the diameter of the system exceeds $\alpha$.
For each $t \leq T_\alpha$, being a connected graph,
$G$ must include an edge $(i,j)$
whose length $|\delta_{ij}(t)|$ exceeds $\alpha/\nu$. 
That edge belongs to $G_t$ with probability $p$;
therefore $\mathbb{E}\, K_{\alpha/\nu} \geq p \,\mathbb{E}\, T_\alpha$;
hence $\mathbb{E}\, T_\alpha \leq \frac{1}{p} \mathbb{E}\, N_{\alpha/\nu}$.

\vspace{0.2cm}
\begin{mdframed}
\begin{theorem}\label{convergence-bound-DMC}
$\!\!\! .\,\,$
The robots align themselves within distance $\epsilon<1$ of a fixed plane  
in expected time $O\bigl(d^2n^4/p^3 \rho^2 \bigr) \log(n/\rho\epsilon)$,
where $d$ is the maximum degree of the underlying communication network,
$n$ is the number of robots, $1-p$ is the probability of edge failure,
and $\rho$ is the smallest communication weight.
\end{theorem}
\end{mdframed}
\vspace{0.2cm}

\vspace{0.4cm}
\begin{figure}[htb]
\begin{center}
\hspace{0cm}
\includegraphics[width=7cm]{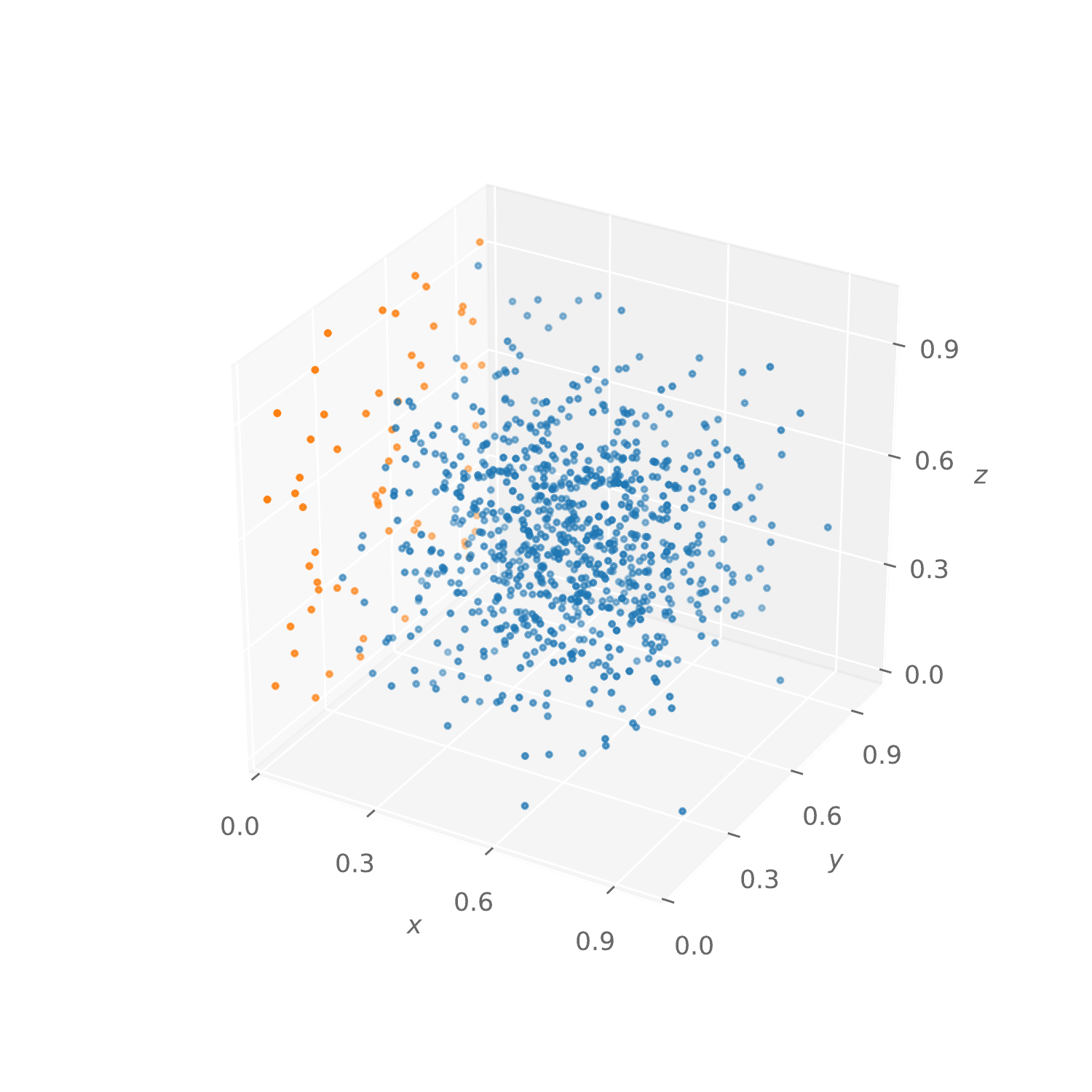}
\hspace{0cm}
\includegraphics[width=7cm]{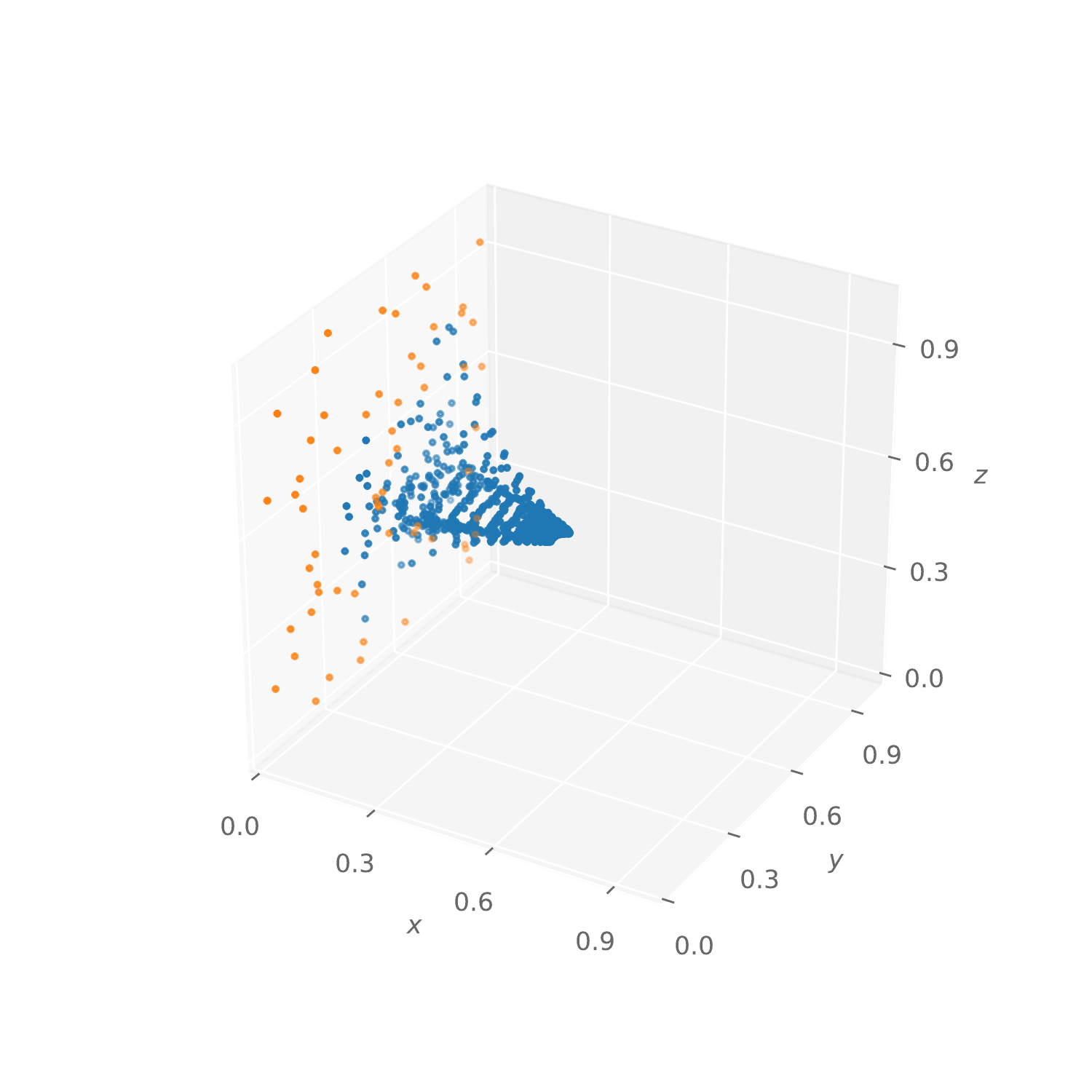}
\end{center}
\vspace{-0.5cm}
\caption{\small 
Simulation of a network of 900 robots in 3D with 60 of them (orange dots) pinned to
a fixed plane $X=0$. The initial positions
of the moving robots (blue dots) are random in the unit cube $[0,1]^3$ (left figure).
The underlying network $G$ is a 30-by-30 grid graph,
with edge failure probability equal to 0.3. The 60 nodes
on two opposite sides of the grid are pinned to $X=0$.
The figure on the right
displays the robots after 200 steps.
\label{fig-robots}}
\end{figure}


\vspace{1cm}


\bigskip\bigskip\bigskip

\newpage
\begin{thebibliography}{1}

{\small

\bibitem{blondelHOT05}
Blondel, V.D., Hendrickx, J.M., Olshevsky, A., Tsitsiklis, J.N.
{\em Convergence in multiagent coordination, consensus, and flocking},
Proc. 44th IEEE Conference on Decision and Control, Seville, Spain, 2005. 

\bibitem{bulloBk} 
Bullo, F., Cort\'es, J., Martinez, S., 
{\rm Distributed Control of Robotic Networks},
Applied Mathematics Series, Princeton University Press, 2009.

\bibitem{chazelle-total}
Chazelle, B.
{\em The total $s$-energy of a multiagent system},
SIAM J. Control Optim. 49 (2011), 1680--1706.

\bibitem{chazelle-Energ2-2019} 
Chazelle, B. 
{\em  A sharp bound on the $s$-energy and its applications to averaging systems},
IEEE Trans. Automatic Control 64 (2019), 4385--4390.

\bibitem{chazFlockPaperI}
Chazelle, B.
{\em The convergence of bird flocking},
J. ACM 61 (2014), 21:1--35.

\bibitem{CuckerSmale1}
Cucker, F., Smale, S.
{\em Emergent behavior in flocks},
IEEE Trans. Automatic Control 52 (2007), 852--862.


\bibitem{HendrickxB}
Hendrickx, J.M., Blondel, V.D.
{\em Convergence of different linear and non-linear 
Vicsek models}, 
Proc. 17th International Symposium on Mathematical 
Theory of Networks and Systems (MTNS2006), Kyoto (Japan), 
July 2006, 1229--1240.

\bibitem{jadbabaieLM03}
Jadbabaie, A., Lin,  J., Morse, A.S.
{\em Coordination of groups of mobile autonomous agents 
using nearest neighbor rules},
IEEE Trans. Automatic Control 48 (2003), 988--1001.

\bibitem{spielman19}
Spielman, D.A.
{\rm Spectral and Algebraic Graph Theory},
Book draft, available at: http://cs-www.cs.yale.edu/homes/spielman/sagt/sagt.pdf, 2019
(page 53--54).

\bibitem{sugihara-suzuki-1990}
Sugihara, K., Suzuki, I.
{\em Distributed Motion Coordination of Multiple Mobile Robots},
Proceedings. 5th IEEE International Symposium on Intelligent Control (1990), pp. 138--143 vol.1, doi: 10.1109/ISIC.1990.128452.

\bibitem{vicsekCBCS95}
Vicsek, T., Czir\'{o}k,  A., Ben-Jacob,  E.,
Cohen,  I., Shochet, O. 
{\em Novel type of phase transition in a system of 
self-driven particles},
Physical Review Letters 75 (1995), 1226--1229.

}
\end{thebibliography}
\end{document}